\documentclass[aip,jcp,twocolumn,10pt]{revtex4-2}
\usepackage{times}
\usepackage{amsmath}
\usepackage{amsfonts}
\usepackage{graphics,graphicx}
\usepackage{color}
\usepackage{tabularx}
\usepackage[T1]{fontenc}

\newcommand{\citen}[1]{\onlinecite{#1}}

\begin{document}  

\date{\today}

\title{The Li-F-H Ternary System at High Pressures}
\author{Tiange Bi}
\affiliation{Department of Chemistry, State University of New York at Buffalo, Buffalo, NY 14260-3000, USA}                              %

\author{Andrew Shamp} 
\affiliation{Department of Chemistry, State University of New York at Buffalo, Buffalo, NY 14260-3000, USA}
\author{Tyson Terpstra}
\affiliation{Department of Chemistry, State University of New York at Buffalo, Buffalo, NY 14260-3000, USA}
\author{Russell J. Hemley}\email{rhemley@uic.edu} 
\affiliation{Departments of Physics and Chemistry, University of Illinois at Chicago, Chicago, Illinois 60607, USA}
\author{Eva Zurek}\email{ezurek@buffalo.edu}
\affiliation{Department of Chemistry, State University of New York at Buffalo, Buffalo, NY 14260-3000, USA}

\begin{abstract}
Evolutionary crystal structure prediction searches have been employed to explore the ternary Li-F-H system at 300~GPa. Metastable phases were uncovered within the static lattice approximation, with LiF$_3$H$_2$, LiF$_2$H, Li$_3$F$_4$H, LiF$_4$H$_4$, Li$_2$F$_3$H and LiF$_3$H lying within 50~meV/atom of the 0~K convex hull. All of these phases contain H$_n$F$^-_{n+1}$ ($n=1,2$) anions, and Li$^+$ cations. Other structural motifs such as LiF slabs, H$_3^+$  molecules and F$^{\delta-}$ ions are present in some of the low enthalpy Li-F-H structures. The bonding within the H$_n$F$^-_{n+1}$ molecules, which may be bent or linear, symmetric or asymmetric, is analyzed. The five phases closest to the hull are insulators, while LiF$_3$H is metallic and predicted to have a vanishingly small superconducting critical temperature. This study lays the foundation for future investigations of the role of temperature and anharmonicity on the stability and properties of compounds and alloys in the Li-F-H ternary system.
\end{abstract}

\maketitle

\section{Introduction}

The incorporation of hydrogen in materials is of both fundamental and technological interest. Hydrogen undergoes compound formation with elements across the periodic table, and can adopt various bonding schemes varying from ionic to covalent, to van der Waals when present in molecular form. Technologically, hydrides have been employed in nuclear reactor applications, but their formation can precipitate corrosion and embrittlement \cite{BANOS2018129}. More recently, a wide variety of hydrides have been considered for energy applications such as for hydrogen storage \cite{LEY2014122} and in battery materials. \cite{Kim:2019a} 

High pressure has been explored as a route to increase the amount of hydrogen that may be stored in materials,  \cite{Song:2013a,Struzhkin:2007a} and to synthesize novel superconductors. \cite{Zurek:2018d}
A broad range of structures can form under pressure, many of which can be very hydrogen rich. Hydrogenic motifs that have been identified include quasimolecular H$_2$ units, atomic H$^-$, as well as extended hydrogenic lattices. \cite{Zurek:2018m,Zurek:2016d,Zurek:2016j} Many of these are very high-temperature superconductors, as first established for H$_3$S  and LaH$_{10}$. These phases were predicted \cite{Duan:2014,liu2017potential,peng2017hydrogen} and synthesized \cite{drozdov2015conventional,Geballe:2018a} under pressure, with measured superconducting critical temperatures, $T_c$s, as high as 203~K \cite{drozdov2015conventional} at 150~GPa, and 250-260~K around 200~GPa,\cite{somayazulu2019evidence,drozdov2019superconductivity} respectively. The theoretical studies of compressed binary hydrides have paved the way to interest in more complex ternary systems including LiPH$_6$, \cite{shao2019ternary} LiP$_2$H$_{14}$,\cite{li2020chemically} Li$_2$MgH$_{16}$, \cite{sun2019route}, CH$_4$ intercalated H$_3$S, \cite{cui2020route,sun2020computational}, SH$_3$-SeH$_3$, \cite{liu2018effect} MgCH$_4$, \cite{tian2015predicted} H$_3$P$_{0.15}$S$_{0.85}$, \cite{fan2016high} CaSH$_3$, \cite{Zurek:2020g} and CaYH$_{12}$ \cite{liang2019potential}, which are predicted to possess $T_c$ values of 100-473~K at pressures ranging from 100-230~GPa. In fact, recently a $T_c$ of 288~K has been reported experimentally in a phase (or phases) in the C-S-H ternary system near 270~GPa. \cite{Snider:2020a} 

The ionic solid LiF has been employed as a radiation dosimeter, in molten salt coolants for nuclear reactors, and it is a test system for models of ionic solids. With its large band gap, LiF has the lowest refractive index of all common infrared materials. It also possesses the highest UV transmission of any material, being able to transmit significantly into the VUV region. The properties of LiF render it important for a wide range of applications in high pressure research, and its properties under static high pressures have been widely studied (see Ref.\ \citen{Kirsch:2019a,ao2009strength}). This includes the use of LiF as a pressure standard and pressure-transmitting medium in diamond anvil cell experiments. \cite{Liu:2007a,Dong:2014a} Moreover, because LiF is observed to remain optically transparent to at least 900~GPa \cite{fratanduono2011refractive} it has been employed extensively as a window material in dynamic compression experiments, \cite{Barker:1972a} e.g. for interferometric velocimetry (VISAR) diagnostics. 
Because of the importance of LiF in these experiments, its high pressure properties continue to be of great current interest. \cite{fratanduono2011refractive,Liujap:2015,Davis:2016a,Myint:2019a,Kirsch:2019a,ao2009strength,wise1986laser,Driver:2017a,Spataru:2015a,LiF1,LiF2,duanjap:2020,He:2012a,knudson2015direct,Celliers:2018a}

The interaction of LiF with hydrogen under pressure has not been explored in detail experimentally or theoretically. Early on it was suggested that LiH$_2$F, where a hydrogen atom or H$_2$ molecule is placed in the pseudo-octahedral holes in the $B1$ structure of LiF,  might be a way to achieve the metallization of hydrogen and concomitant superconductivity as discussed above. \cite{Gilman:1971a} Also, possible pressure-induced chemical reactions between hydrogen and LiF windows in dynamic compression experiments \cite{knudson2015direct,Celliers:2018a} could affect the interpretation of the results. Given the above, and the propensity for formation of stable and metastable hydrides under pressure, we were motivated to apply crystal structure prediction techniques to the Li-F-H ternary system to multimegabar pressure. By now the phase diagrams of the elemental (Li, \cite{Li} H$_2$, \cite{H} and F$_2$ \cite{F}) and binary (LiH$_n$, \cite{LiH,LiH2-H6} H$_n$F, \cite{HF1,HF2} and LiF \cite{LiF1,LiF2}) systems have been studied computationally at 300~GPa, providing the basis for us to explore the high pressure phases of the ternary system. 

Using evolutionary crystal structure prediction techniques, we find a number of metastable phases that are within 50~meV/atom of the convex hull and are therefore potentially synthesizable. Common structural motifs present in these phases include  H$_n$F$_{n+1}^-$ anions of various lengths and Li$^+$ counter-cations. Most of these crystalline lattices are calculated to be wide-gap insulators, with the exception of LiF$_3$H, which is predicted to be metallic and superconducting below 0.1~K.

\section{Computational Details}

Crystal structure prediction searches were performed using density functional theory (DFT) coupled with the evolutionary algorithm (EA), \textsc{XtalOpt}, release 9. \cite{Zurek:2011a,Falls:2016,Zurek:2020i} Duplicate structures were detected via the \textsc{XtalComp} algorithm \cite{Lonie:2012}. EA runs were carried out on LiFH$_n$ ($n$ = 2, 3), LiF$_2$H$_n$ ($n$ = 1-2), LiF$_3$H$_n$ ($n$ = 1-3), LiF$_4$H$_n$ ($n$ = 1, 4), Li$_2$F$_m$H ($m$ = 1-3), Li$_2$FH$_2$, Li$_3$F$_m$H ($m$ = 1-4), Li$_4$FH$_4$, Li$_4$F$_3$H, and Li$_4$F$_4$H stoichiometries at 300~GPa employing 2-3 formula units (FU) within the simulation cells. The lowest enthalpy structures obtained from each search were relaxed at 300~GPa. 

Geometry optimizations and electronic structure calculations were performed using DFT as implemented in the Vienna \textit{Ab-Initio} Simulation Package (VASP) version 5.4.1, \cite{VASP5.4.1} with the gradient-corrected exchange and correlation functional of Perdew-Burke-Ernzerhof (PBE). \cite{Perdew:1996} The projector augmented wave (PAW) method \cite{Bloch:1994} was used to treat the core states, along with a  plane-wave basis set with an energy cutoff of 700~eV. The F 2$s^2$2$p^5$, H 1$s^1$ and Li 1$s^2$2$s^1$ electrons were treated explicitly using the PAW-PBE F, PAW-PBE H and PAW-PBE Li\textunderscore sv POTCARs.  
In the EA searches only the  Li 2$s^1$ electrons were considered as valence, because exploratory EA runs for LiF$_4$H$_3$ with the two sets of POTCARs found the same lowest enthalpy structures, and their volumes differed by less than 0.5\%. The $k$-point grids were generated using the $\Gamma$-centered Monkhorst-Pack scheme, and the number of divisions along each reciprocal lattice vector was chosen such that the product of this number with the real lattice constant was 30~\AA{} in the structure searches, and 50~\AA{} otherwise. The crystal orbital Hamiltonian
populations (COHPs), \cite{dronskowski1993crystal} and the negative of the COHPs integrated to the Fermi level ($-$iCOHPs) were calculated using the LOBSTER package \cite{maintz2013analytic}.

Phonon calculations were performed using VASP combined with the Phonopy\cite{Phonopy} package under the harmonic approximation. The supercells were chosen such that the number of atoms within them was always greater than 100. Infrared (IR) spectra were simulated using the Phonopy-Spectroscopy package. \cite{IR1,IR2} Density functional perturbation theory, DFPT, as implemented in the Quantum Espresso (QE) \cite{giannozzi2009quantum} program, was used to obtain the dynamical matrix of the LiF$_3$H phase. The Li, F, and H pseudopotentials, obtained from the GRBV pseudopotential library, adopted used the  1$s^2$2$s^{0.55}$, 2$s^2$2$^5$, and 1$s^1$ valence configurations, respectively. The plane-wave basis set cutoff energies were set to 90~Ry, and the Brillouin-zone sampling scheme of Methfessel-Paxton \cite{methfessel1989high} using a smearing of 0.002~Ry and a 12 $\times$ 12 $\times$ 12 $k$-point grid was employed. The electron phonon coupling (EPC), $\lambda$, was calculated using a set of Gaussian broadenings of 0.012~Ry, and a 3 $\times$ 3 $\times$ 3 $q$-grid. 

The geometry optimization of the (F-H-F)$^-$ anion in the gas phase was performed using the Amsterdam Density Functional (ADF) software package. \cite{ADF,ADF2} The basis functions consisted of an all electron triple-$\zeta$ Slater-type basis set with polarization functions (TZP) from the ADF basis set
library, \cite{van2003optimized} and the PBE \cite{Perdew:1996} functional was employed.

\section{Results}
\subsection{Thermodynamic Stability}
 
At 300~GPa and 0~K, the thermodynamically stable phases located on the three-dimensional convex hull are the elemental and binary phases: Li, H$_2$, F$_2$, LiF, HF, LiH, LiH$_2$, and LiH$_6$. However, all of the ternary Li-F-H phases uncovered here are dynamically stable. Among them LiF$_3$H$_2$, LiF$_2$H, Li$_3$F$_4$H, LiF$_4$H$_4$, and Li$_2$F$_3$H are within 41~meV/atom of the hull  (Figure S3, Table S1), while LiF$_3$H$_3$, Li$_3$FH, LiF$_3$H, Li$_4$F$_3$H, Li$_4$FH$_4$, LiF$_4$H, Li$_2$FH, Li$_2$F$_2$H, and Li$_3$F$_2$H are within 58-91~meV/atom of the hull. All other phases are less stable with the most unfavorable candidate that was identified, LiFH$_3$, lying 222.5~meV/atom above the hull. As described in Figure S5, the LiH$_2$F phase found in the EA searches was considerably more stable than those proposed previously by Gilman. \cite{Gilman:1971a}

When the zero-point-energy (ZPE) was taken into consideration, LiF$_3$H joined the aforementioned five phases  to lie less than 50~meV/atom from the hull (Figure S4, Table S2). The reaction enthalpies, $\Delta H_F$, of the decomposition of LiF$_3$H$_2$, LiF$_2$H, Li$_3$F$_4$H, and Li$_2$F$_3$H into LiF and HF were computed to be -13.3 (10.9), -29.3 (-28.5), -33.1 (-33.8), and -40.5 (-41.1)~meV/atom when the ZPE was neglected (taken into consideration), consistent with the fact that they lie above the convex hull. Other reactions that thermodynamically destabilize two additional phases of interest and their computed reaction enthalpies are: $\text{LiF}_4\text{H}_4 \rightarrow \text{LiF} +3\text{HF}+\frac{1}{2}\text{H}_2$ ($\Delta H_F =$ -33.4 (-37.0)~meV/atom), and $\text{LiF}_3\text{H} \rightarrow \text{LiF}+ \text{HF}+\frac{1}{2}\text{F}_2$ ($\Delta H_F =$ -62.5 (-49.0)~meV/atom). 

\begin{figure*}
\begin{center}
\includegraphics[width=1.8\columnwidth]{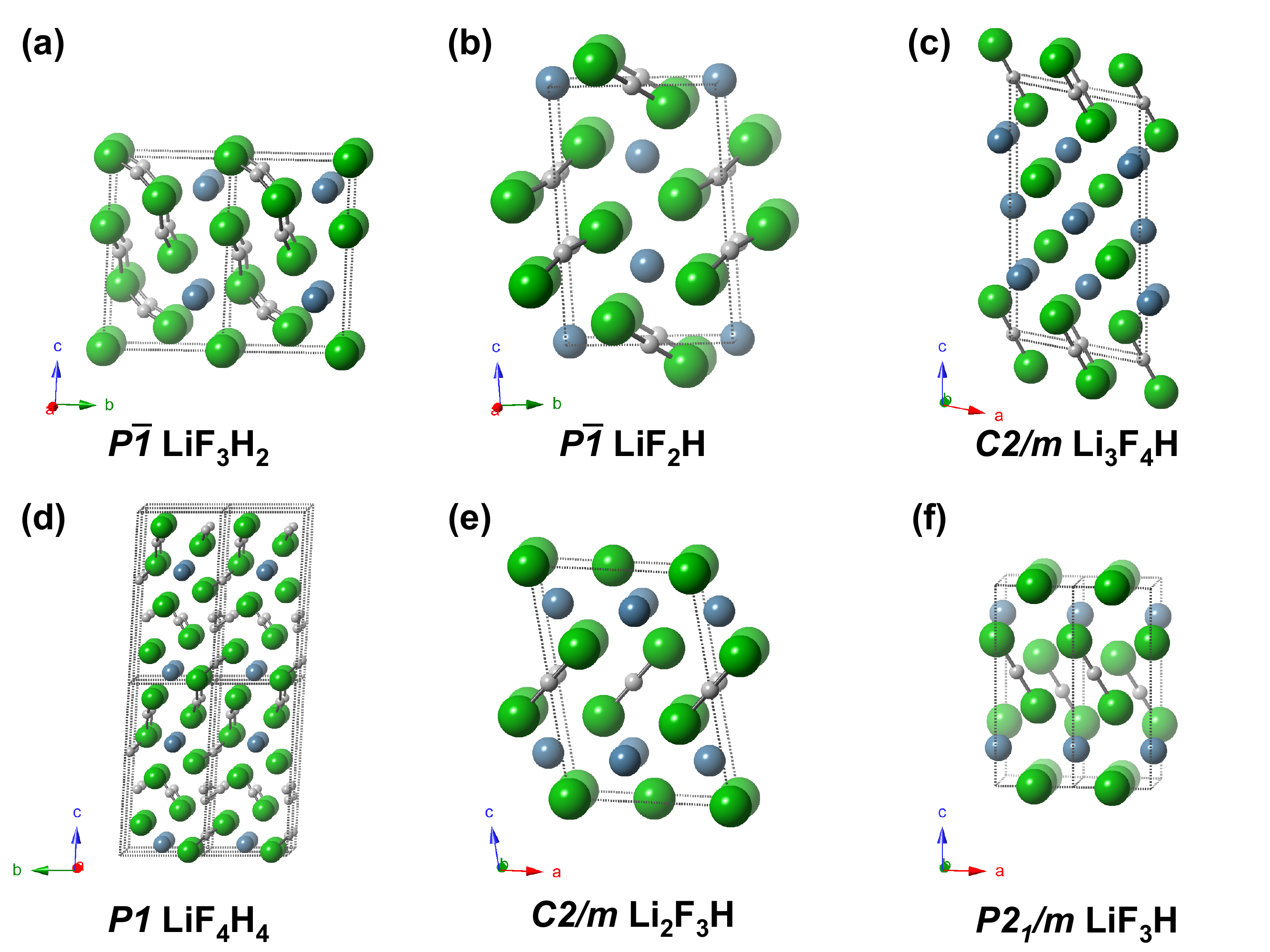}
\end{center}
\caption{Crystal structures of predicted Li-F-H phases at 300~GPa: (a) $P\overline{1}$ LiF$_3$H$_2$, (b) $P\overline{1}$ LiF$_2$H, (c)
$C2/m$ Li$_3$F$_4$H, (d) $P1$ Li$_4$H$_4$, (e) $C2/m$ Li$_2$F$_3$H, and (f) $P2_1/m$ LiF$_3$H. Li/F/H atoms are
colored blue/green/white.}  
\label{fig:structure}
\end{figure*} 

A recent data-mining study found that the 90$^{th}$ percentile of the 0~K DFT-calculated metastability of all of the compounds within the Inorganic Crystal Structure Database (ICSD) was $\sim$70~meV/atom \cite{materialsproject}. For matter under pressure the choice of the synthetic pathway, including the precursors and temperatures employed, might be able to produce kinetically stable phases. Several examples of metastable hydrides that have been synthesized under pressure include C$_x$S$_y$H$_z$, \cite{Snider:2020a,Zurek:2020b,Sun:2020a} PH$_n$, \cite{drozdov2015superconductivity,Zurek:2015j,Flores-Livas:2016,Zurek:2017c} and CaH$_{2.5}$. \cite{Zurek:2018b} Moreover, anharmonic and quantum nuclear effects, not considered here, can be important in determining the stability of compounds, in particular those containing light elements.  \cite{errea2020quantum} Therefore, we chose to analyze the structure, bonding and properties of phases that were within 50~meV/atom of the convex hull at 0~K.

\subsection{Structural Properties}
Our first-principles based crystal structure prediction searches predicted that $P\overline{1}$ LiF$_3$H$_2$, shown in Figure \ref{fig:structure}(a), is the closest phase to the 300~GPa convex hull.  A prominent structural motif within LiF$_3$H$_2$ is the V-shaped (FH-F-HF)$^-$ anion, whose H-F$_{\text{middle}}$-H bond angle measures  110.0$^\circ$, and F$_\text{terminal}$-H-F$_\text{middle}$ bond angles measure 160.3$^\circ$ and 165.7$^\circ$. The F$_\text{terminal}$-H bonds comprising this building block are shorter than the H-F$_\text{middle}$ (0.998 and 1.004~\AA{} vs.\ 1.107 and 1.099~\AA{}, respectively). Each lithium cation is coordinated to six fluorine atoms at distances ranging from 1.587-1.638~\AA{}, and each fluorine atom is coordinated to two lithium cations. One of the Li-F$_\text{terminal}$ contacts in each  (FH-F-HF)$^-$ unit measures 1.587~\AA{}, and the fluorine involved in this contact forms the shortest H-F$_\text{terminal}$ bond with the largest F$_\text{terminal}$-H-F$_\text{middle}$ bond angle. 

The phase that is second closest to the convex hull, $P\overline{1}$ LiF$_2$H, shown in Figure \ref{fig:structure}(b), resembles LiF$_3$H$_2$, but the (FH-F-HF)$^{-}$ units are replaced by bifluoride, (F-H-F)$^{-}$, anions. This phase is characterized by layers containing Li$^+$ and (F-H-F)$^{-}$ units that are stacked in an ABCABC... fashion along the $c$-axis. The (F-H-F)$^{-}$ molecules in layers A and B are bent with a bond angle of 165.3$^\circ$, and bond lengths of 1.034 and 1.067~\AA{}. In going from one layer to the next, the orientation of the (F-H-F)$^{-}$ units change via a rotation of 180$^\circ$ around the $c$-axis. In layer C, the (F-H-F)$^{-}$ anion is linear and symmetric, with both of the H-F bonds measuring 1.037~\AA{}.

The gas phase bifluoride anion possesses $D_{\infty h}$ symmetry and is characterized by a strong three-center four-electron bond (3c-4e). \cite{Pimentel:1951a} In crystals the anion is often bent with two unequal F-H bond lengths, and in some cases may better be described as (F$^-$)(HF) with one formal H-F two-center two-electron (2c-2e) bond and one F$^-\cdots$H hydrogen bond. \cite{grabowski2016fhf,Bernal:2014a} A deformation from the ideal symmetry is often a result of the interaction with Lewis acids in the crystal lattice, or with external hydrogen bonds. Although less common, an (FHF)$^-\cdots$HF structural motif, which may be treated as an F$_3$H$_2^-$ ion when the bonds are nearly symmetric has also been observed in various salts. \cite{Manson:2007a,Troyanov:2005a,Mootz:1986a} A few inorganic compounds containing other suprahalide anions with the general formula H$_n$F$_{n+1}^-$ and $n=3-6$ are also known. \cite{Bernal:2014a,Wiechert:1998a}  

The two molecular motifs present within $P\bar{1}$ LiF$_3$H$_2$ and $P\bar{1}$ LiF$_3$H$_2$, (FH-F-HF)$^-$ and (F-H-F)$^{-}$, are also found in the other four compounds that are calculated to lie within 50~meV/atom of the 300~GPa convex hull (Figures \ref{fig:structure}(c-f)). Bifluoride anions are present in all of the remaining structures, and LiF$_4$H$_4$ also contains the longer five-atom chain. The smaller H$_n$F$^-_{n+1}$ anion is the only molecular species found in phases where the H:Li ratio $\le$~1 and the H:F ratio $\le$~0.5.

We now examine the structural peculiarity of these phases, starting with those that only contain the bifluoride anion. The $C2/m$ symmetry Li$_3$F$_4$H phase consists of layers of bifluoride anions that lie parallel to each other stacked along the $c$-axis. These are separated by a five atom thick LiF slab made from alternating Li and F layers, with Li comprising both the top and bottom layers.  At ambient conditions LiF crystallizes in the $B1$ ($Fm\bar{3}m$) structure, and first-principles calculations have shown that it does not undergo a phase transition to the $B2$ structure to at least 1~TPa at 0~K. \cite{LiF1} Not surprisingly, the LiF layer in Li$_3$F$_4$H can be described as a slightly distorted slab cut out from the $B1$ phase. The segregation into distinct layers observed within Li$_3$F$_4$H is consistent with the finding that its decomposition into LiF and HF is preferred.  

The slightly less stable $C2/m$ symmetry Li$_2$F$_3$H phase can be constructed from Li$_3$F$_4$H by removing two layers (one of lithium and one of fluorine) from the LiF slab. In both of these compounds the (F-H-F)$^-$ units are linear and symmetric with H-F bond lengths of 1.041~\AA{}, and the fluorine atom is coordinated to three lithium ions in the LiF slab.  $P2_1/m$  LiF$_3$H also contains layers of linear (F-H-F)$^-$ anions, however they are not symmetric with H-F bond lengths of 1.079 and 1.014~\AA{}, and each one of the fluorine atoms within them are coordinated to two lithium ions in the outer most layer of the LiF slab. It is conceivable that phases with these same stoichiometries, but different numbers of atoms in the unit cell, or different stoichiometries that are composed of a single layer of bifluoride anions separated by LiF layers of varying thickness may also be metastable, with enthalpies of formation falling within the realm of synthesizability. 

The most complex structure identified, $P1$ LiF$_4$H$_4$, contains both the short (F-H-F)$^{-}$ and the long (FH-F-HF)$^-$ anions, along with triangular H$_3^+$ cations. This compound can also be viewed as a layered phase, with  bifluoride anions and H$_3^+$ cations comprising one set of layers. Each one of the fluorine atoms in these  (F-H-F)$^-$ units is coordinated to a single lithium ion within the next layer, which also contains (FH-F-HF)$^-$ motifs wherein both of the terminal fluorines are coordinated to two lithium ions. Thus, each Li$^+$ is coordinated to seven fluorines, three belonging to  (F-H-F)$^-$ and four to  (FH-F-HF)$^-$, at distances ranging from 1.601 to 1.662~\AA{}.

The bifluoride anion linkages in LiF$_4$H$_4$ are not linear, with a bond angle of 175.5$^\circ$, nor are they symmetric, with bond lengths measuring 1.035 and 1.042~\AA{}. The H$_3^+$ molecular units resemble the trihydrogen cation -- one of the most abundant interstellar molecules \cite{Oka12235}, which assumes an equilateral triangle and forms a three-center two-electron (3c-2e) bond. However, the  H$_3^+$ molecule found here is not quite a perfect equilateral triangle, with bond angles of 58.4, 60.4 and 61.2$^\circ$, and bond lengths of 0.805, 0.782, and 0.799~\AA{}. This structural motif has been predicted to exist in H$_5$Cl, \cite{Wang:H5Cl,Zeng:H5Cl,Duan:H5Cl} as well as H$_2$F, H$_3$F and H$_5$F \cite{Duan:H5Cl} phases under pressure. The geometric parameters of the (FH-F-HF)$^-$ molecules resemble those in the LiF$_3$H$_2$ phase, with bond angles of 107.9$^\circ$ (H-F$_\text{middle}$-H), as well as 160.1$^\circ$ and 166.4$^\circ$ (F$_\text{terminal}$-H-F$_\text{middle}$), and bond lengths of 0.997 and 1.002~\AA{} (H-F$_\text{terminal}$), as well as 1.110 and 1.098~\AA{} (H-F$_\text{middle}$). Similar to what was found in $P\bar{1}$ LiF$_3$H$_2$, the terminal fluorine atom that is closest to a lithium ion has the smaller F-H-F bond angle, and the shorter H-F bond length. 

\subsection{Bonding}

Because of the difficulties inherent in measuring the position of hydrogen atoms in crystals, the geometry of the bifluoride anion in simple salts such as KHF$_2$ \cite{Ibers:1964} and NaHF$_2$ \cite{McGaw:1963} has been debated. Two types of bonding scenarios are possible: a delocalized 3c-4e bond in the symmetric structure, vs.\ one 2c-2e bond and one extremely strong hydrogen bond in the unsymmetric geometry. In more complex salts hydrogen bonding with other moieties, or electrostatic interactions with strong Lewis acids can lead to a deviation from linearity. \cite{grabowski2016fhf} As discussed above, examples of both of these scenarios are present within the phases predicted here. 

In our discussion it is also instructive to consider the evolution of the low temperature phases of HF under pressure. At ambient conditions the $Cmc2_1$ symmetry HF phase contains planar zigzag chains of hydrogen-bonded molecules that are held together via van der Waals forces. \cite{Johnson:1975a} Between 25-143~GPa a $Cmcm$ symmetry structure where each fluorine atom is symmetrically bonded to two hydrogen atoms is preferred. \cite{Pinnick:1989a,HF2} First principles calculations have predicted a further transition to  a $Pnma$ symmetry structure with nearly symmetric H-F \cite{HF2} bonds that is stable until at least 900~GPa. \cite{WANG2020}

In order to better understand the bonding within the (F-H-F)$^-$  and (FH-F-HF)$^-$ anions we calculated the negative of the crystal orbital Hamilton populations between select atoms integrated to the Fermi level (-iCOHP) because they can be used to gauge the bond strength. The results obtained for the Li-F-H phases can be compared with values calculated for a symmetric (F-H-F)$^-$ molecule optimized in the gas phase at 0~GPa, and $Pnma$ HF optimized at 300~GPa (Table\ \ref{tab:bonds}).
 
  \begin{table}[b!]
\centering
\caption{Distances between fluorine and hydrogen atoms, and F-H-F angles within the bifluoride anion in the gas phase, as well as $Pnma$ HF at 300~GPa, and their corresponding crystal orbital Hamilton populations integrated to the Fermi level (-iCOHP). These same quantities are given for the bifluoride ions present within the Li-F-H phases at 300~GPa. } 
\begin{tabular}{  c  c  c  c  c  c   } 
\hline
\hline
System  & \multicolumn{2}{c}{Distance (\AA{})} &  Bond Angles ($^\circ$) & \multicolumn{2}{c}{$-$iCOHP (eV/bond)} \\

             & F1-H & F2-H    & F-H-F & F1-H & F2-H  \\ 
\hline 
(F-H-F)$^-$ &  1.160 & 1.160 & 180.0  & 3.61 & 3.61 \\
HF       & 1.047 & 1.048 & 175.3 & 7.46 & 7.09 \\

LiF$_2$H$^a$     & 1.037 & 1.037 & 180.0 & 6.89 & 6.89  \\
LiF$_2$H$^a$ & 1.034 & 1.067 & 165.3 & 7.10 & 6.33  \\
Li$_3$F$_4$H & 1.041 & 1.041 & 180.0 & 6.95 & 6.95  \\
LiF$_4$H$_4$ & 1.035 & 1.042 & 175.5 & 7.33 & 7.06 \\
Li$_2$F$_3$H & 1.041 & 1.041 & 180.0 & 6.96 & 6.96  \\
LiF$_3$H     & 1.014 & 1.079 & 180.0 & 7.81 & 6.22  \\
\hline
\hline
\end{tabular} \\
$^a$ Linear and bent bifluoride anions are present in this phase.
\label{tab:bonds}
\end{table}
 
\begin{figure}
\begin{center}
\includegraphics[width=0.6\columnwidth]{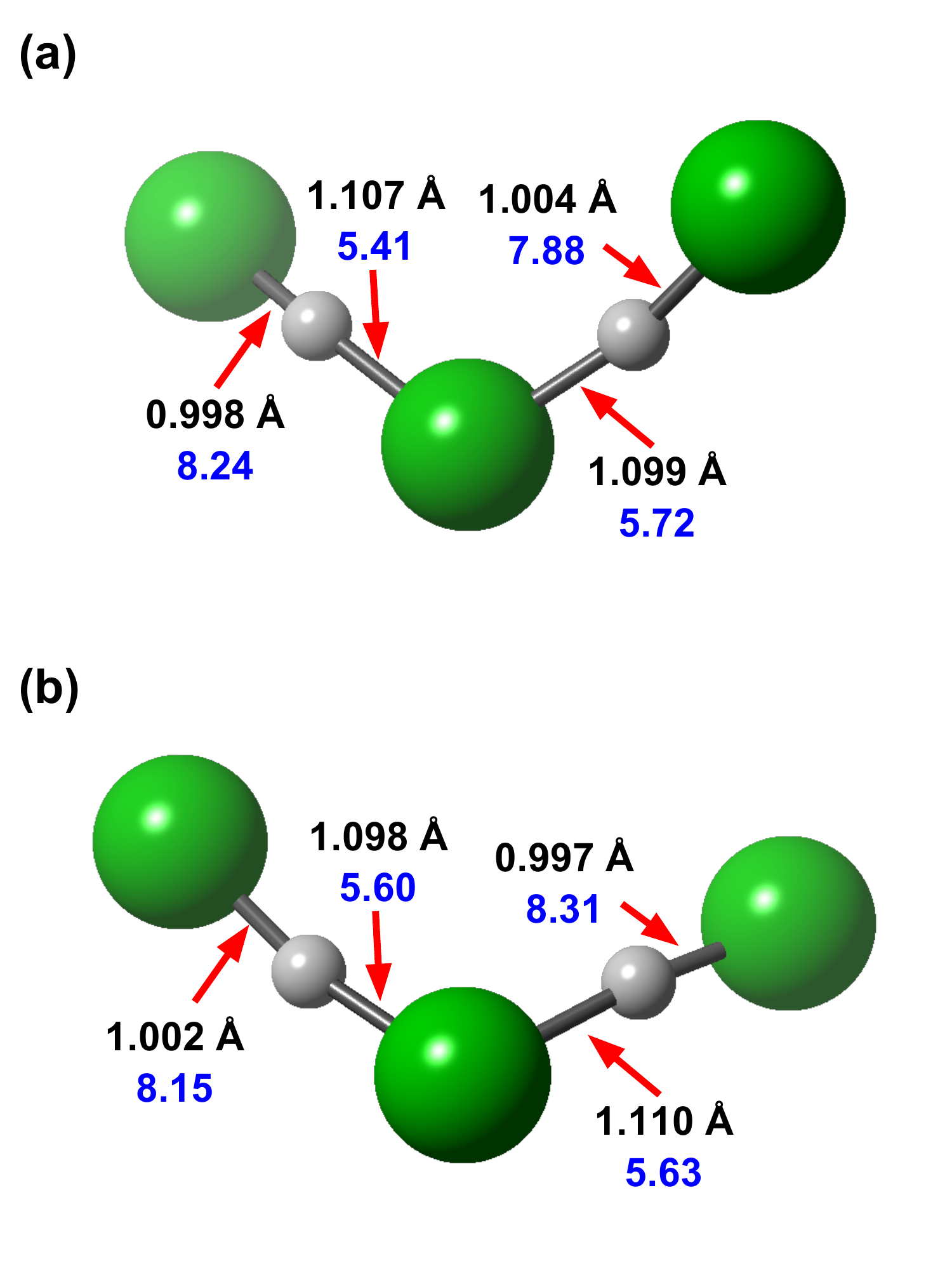}
\end{center}
\caption{The molecular conformation of the (FH-F-HF)$^-$ anions taken from (a) LiF$_3$H$_2$ and (b) LiF$_4$H$_4$ at 300~GPa. The values given in black are the bond lengths in \AA{}, and the values colored in blue are the $-$iCOHPs in eV/bond.}  
\label{fig:molecule}
\end{figure} 

LiF$_2$H, Li$_3$F$_4$H and Li$_2$F$_3$H contain linear bifluoride anions with equivalent F-H bond lengths, and therefore they can be considered 3c-4e bonded species. Because the neighboring Li$^+$ cations are arranged symmetrically about the (F-H-F)$^-$ anions, the geometry of these molecular motifs are not distorted via electrostatic interactions. The F-H bond lengths in these species are shorter than in the bifluoride anion in the gas phase, and the bond strengths, as measured by the -iCOHPs, are therefore almost twice as large, in-line with the bond strengthening that is expected to occur because of increased orbital overlap under pressure. \cite{Zurek:2019k} The calculated F-H bond strengths and lengths in the Li-F-H ternaries are not too different from those found in solid HF at the same pressure. In addition to the $D_{\infty h}$ symmetry (F-H-F)$^-$ unit, LiF$_2$H also contains a bifluoride anion that is bent and asymmetric because of electrostatic interactions with neighboring Li$^+$ cations and other (F-H-F)$^-$ species. Both sets of H-F bonds are stronger and shorter than in the gas phase molecule. Asymmetric (F-H-F)$^-$ also comprise LiF$_4$H$_4$, whose bifluoride anion is slightly bent, and LiF$_3$H, where it remains linear. The bond strengths in the ternaries typically increase with decreasing bond lengths.

Although the asymmetric H-F bonds within the 1D zigzag chains in HF do not have a significantly different bond length, their different chemical environments affect their strengths with calculated -iCOHPs of  7.46 and 7.09~eV/bond.  Generally speaking the H-F bond distances in $Pnma$ HF are slightly larger than in the ternaries, but the -iCOHPs are also larger. This suggests that the complicated chemical environment in the ternary phases with the presence of Lewis acid species destabilizes the bifluoride anions.

Figure \ref{fig:molecule} illustrates the H-F bond lengths and associated $-$iCOHPs in the more complicated (FH-F-HF)$^-$ units that are present in LiF$_3$H$_2$ and LiF$_4$H$_4$. In both cases the H-F$_\text{terminal}$ bonds are shorter and stronger than in HF or in any of the ternaries, whereas the H-F$_\text{middle}$ are longer and weaker ($\sim$1~\AA{} and 7.9-8.3~eV/bond vs.\ $\sim$1.1~\AA{} and 5.4-5.7~eV/bond). This coincides with the 1~atm picture where the  (FH-F-HF)$^-$ motifs are viewed as being composed of two HF units hydrogen bonded to a central F$^-$. \cite{grabowski2016fhf}

Bader charges were calculated to verify the formal charges assigned. The results, provided in Table S3, show that in all of the Li-F-H phases the Bader charge on Li fell between +0.83 and +0.86, in-line with the +1 oxidation state. The charges on the fluorine and hydrogen atoms within the H$_n$F$^-_{n+1}$ anions typically ranged from -0.74 to -0.80, and +0.73 to +0.76, respectively, which is in-line with the values computed for HF at the same pressure, -/+0.75. Generally speaking, the overall charges on the H$_n$F$^-_{n+1}$ anions fell between -0.76 to -0.85, and the charges on the fluorine atoms comprising the LiF slabs were somewhat more negative than in the molecular motifs, -0.82/0.83. Only one phase, LiF$_3$H, deviated from these trends. The reason for this turns out to be key  for the metallicity of this phase, as described below.

\subsection{Electronic Structure and Properties}

\begin{figure*}
\begin{center}
\includegraphics[width=1.2\columnwidth]{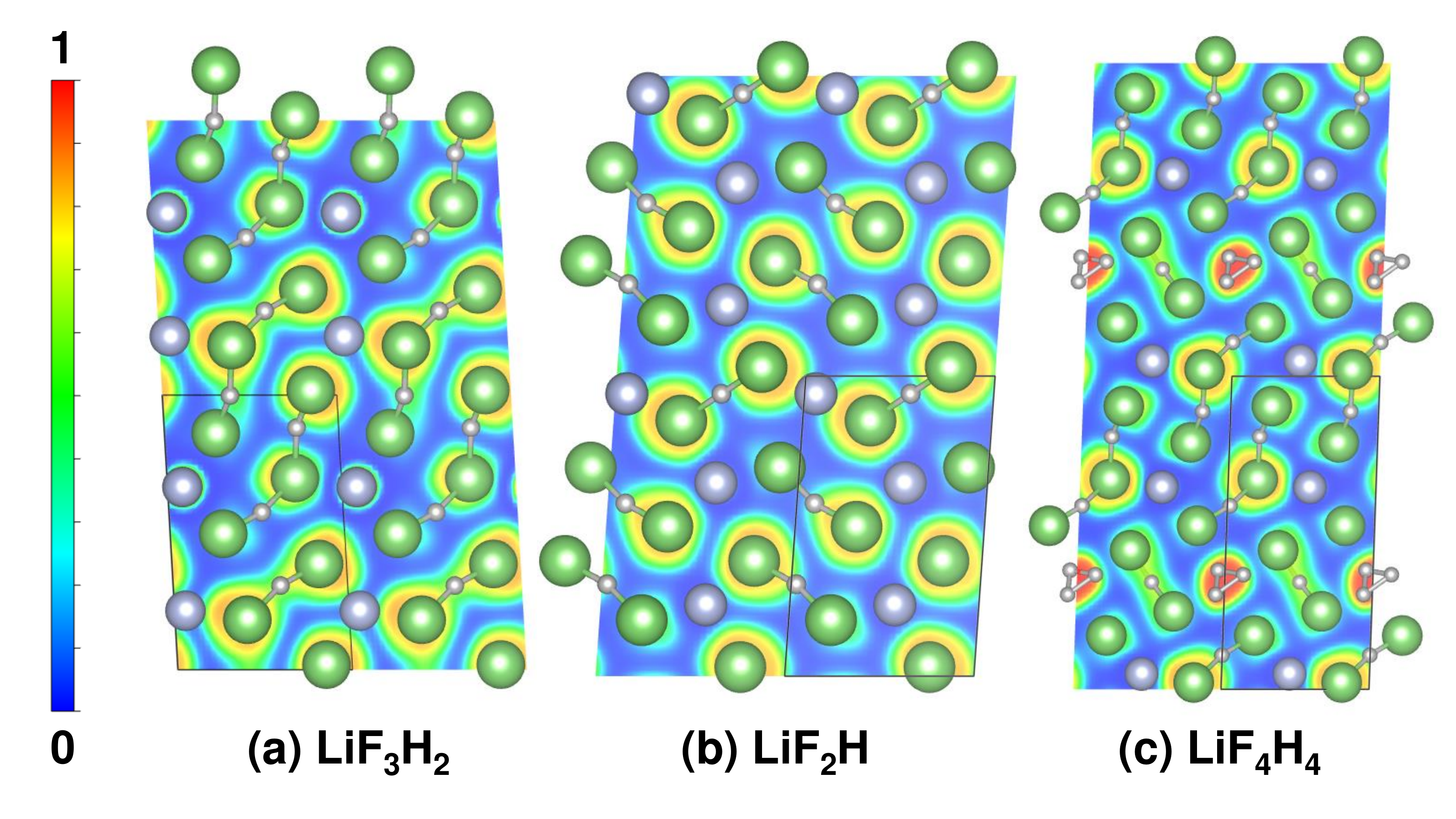}
\end{center}
\caption{Contour maps of the electron localization function (ELF) at 300~GPa shown for 2  $\times$ 2 $\times$ 2 supercells of (a) $P\overline{1}$ LiF$_3$H$_2$ (100) plane,
(b) $P\overline{1}$ LiF$_2$H (001) plane, and (c) $P1$ LiF$_4$H$_4$ (100) plane. Li/F/H atoms are colored blue/green/white.}  
\label{fig:elf}
\end{figure*}

\begin{figure*}
\begin{center}
\includegraphics[width=1.9\columnwidth]{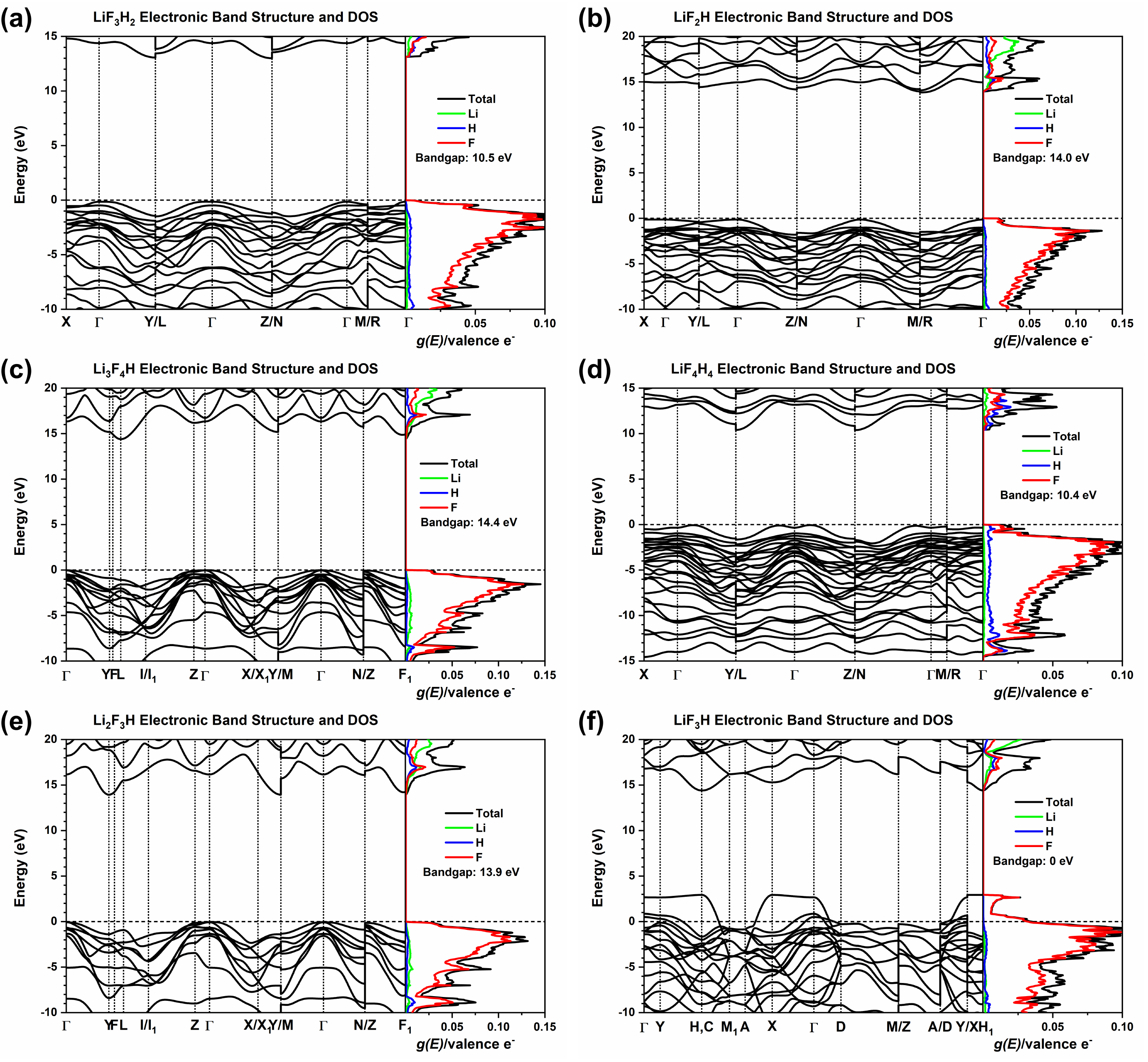}
\end{center}
\caption{Electronic band structure and projected densities of states (PDOS) of (a) $P\overline{1}$ LiF$_3$H$_2$, (b) $P\overline{1}$ LiF$_2$H, (c)
$C2/m$ Li$_3$F$_4$H, (d) $P1$ LiF$_4$H$_4$, (e) $C2/m$ Li$_2$F$_3$H, and (f) $P2_1/m$ LiF$_3$H at 300~GPa. In (a-e) the top of the valence band is set to 0~eV, whereas in (f) the Fermi level is set to 0~eV.}  
\label{fig:bsdos}
\end{figure*}

Figure \ref{fig:elf} plots the electron localization functions (ELFs) of  LiF$_3$H$_2$, LiF$_2$H, and LiF$_4$H$_4$ because they are representative of all of the structural motifs observed. LiF$_3$H$_2$ and LiF$_4$H$_4$ contain the (FH-F-HF)$^-$ anion; LiF$_4$H$_4$ also accommodates the H$_3^+$ unit and asymmetric (F-H-F)$^-$; and  LiF$_2$H contains both symmetric and asymmetric (F-H-F)$^-$. Because the planes of the contours chosen do not necessarily coincide with planes passing through the bifluoride anions and H$_3^+$, and because a single plane cannot pass through (FH-F-HF)$^-$, some discontinuities are present in the ELF contour maps. Therefore, isosurface plots are also provided in Figures\ S6-8. Nonetheless, the ELF plots clearly show high regions of localization around the fluorine atoms, with slightly more density localized along the F-H contacts that are shorter as compared to those that are longer. Moreover, the ELF is high around the H$_3^+$ molecule because of its 3c-2e bond.

The electronic band structure and densities of states (DOS) plots of LiF$_3$H$_2$, LiF$_2$H, Li$_3$F$_4$H, LiF$_4$H$_4$, Li$_2$F$_3$H, and LiF$_3$H at 300~GPa are provided in Figure\ \ref{fig:bsdos}. As expected, the F $p$ orbitals contribute the most to the occupied DOS. The bottom of the conduction band, on the other hand, contains almost equal contributions from F 2$p$, H 1$s$, and Li 2$s$ states, which is consistent with the Bader charges indicating that both H and Li transfer electrons to the most electronegative element present. Except for LiF$_3$H, all of the Li-F-H phases considered here have large band gaps, which are listed in the DOS plots, ranging from 10.4 to 14.4~eV within the PBE functional. The band gap for the $B1$ LiF phase calculated at this level of theory is 15.9~eV at 300~GPa, suggesting that the presence of hydrogen in the ternary phases reduces the band gap slightly. The large band gaps of these ternary hydrides suggest that they remain transparent at 300~GPa. 

Only one phase, $P2_1/m$ LiF$_3$H, was calculated to be metallic at 300~GPa. The DOS at the Fermi level, $E_F$, is almost completely due to the F 2$p$ orbitals. The metallicity of this phase is not a result of pressure-induced band broadening, but rather a result of the dissociation of F$_2$ molecules and the electron count in the compound. Between 70-2500~GPa diatomic fluorine is predicted to adopt the $Cmca$ structure, followed by a transformation to a novel metallic phase with $P4_2/mmc$ symmetry. \cite{F} In contrast, experiments have shown that iodine becomes monoatomic, metallic and superconducting with a $T_c$ of 1.2~K already at 28~GPa. \cite{riggleman1963approach,kenichi2003modulated,shimizu1994pressure} In the full ionic picture the formula of $P2_1/m$ LiF$_3$H can be written as Li$^+$(FHF$^-$)F$^0$. This phase could be insulating if fluorine were present as a diatomic molecule. However, the Bader charges yield Li$^{+0.85}$(FHF$^{-0.52}$)F$^{-0.33}$, and the nearest neighbor distance between two fluorine atoms in the LiF slab separating the bifluoride layers measures 1.610~\AA{}, as compared to the F-F bond length of 1.393~\AA{} in $Cmca$ F$_2$ at the same pressure. Since F$_2$ molecules are not found in $P2_1/m$ LiF$_3$H, an extra electron per formula unit would be required to render it a wide-gap insulator. 

Because $P2_1/m$ LiF$_3$H is metallic, we calculated its electron-phonon coupling (EPC) parameter, $\lambda$, and estimated its $T_c$ using the Allen-Dynes modified McMillan equation \cite{allen1975transition}, $T_c = \frac{\omega_{\text{log}}}{1.2}\exp\left[-\frac{1.04(1+\lambda)}{\lambda-\mu^*(1+0.62\lambda)}\right]$, where $\omega_{\text{log}}$ is the logarithmic average frequency, and $\mu^*$ is the Coulomb pseudopotential, often assumed to be between $\sim$0.1-0.13. The EPC parameter was calculated to be 0.27  and $\omega_{\text{log}}$ was 695.2~K yielding an estimated $T_c$ of 0.092-0.007~K. Thus, in contrast with Gilman's prediction \cite{Gilman:1971a} none of the low-lying metastable Li-F-H ternary phases at 300~GPa are good candidates for pressure-induced hydride based superconductors. Finally, to aid future characterization of these phases, should they ever be synthesized, we present calculated infrared spectra for HF, LiF$_3$H$_2$, LiF$_2$H, Li$_3$F$_4$H, LiF$_4$H$_4$, Li$_2$F$_3$H, and LiF$_3$H in Figures.\ S29-35.

\section{Conclusion}
Using evolutionary crystal structure searches we predict a number of metastable phases, LiF$_3$H$_2$, LiF$_2$H, Li$_3$F$_4$H, LiF$_4$H$_4$, Li$_2$F$_3$H and LiF$_3$H that are within 50~meV/atom of the ternary convex hull within the static-lattice approximation. These potentially synthesizable phases contain the bifluoride anion, (F-H-F)$^-$, or the longer (FH-F-HF)$^-$ molecular motif, along with Li$^+$ cations. LiF$_4$H$_4$ additionally features an H$_3^+$ counter-cation, and LiF slabs are present within Li$_3$F$_4$H. The bonding within the H$_n$F$_{n+1}^-$ motifs encountered here is analyzed, and it ranges from multi-centered interactions to those containing classic two-centered two-electron bonds and a H$\cdots$F$^-$ hydrogen bonds. With the exception of LiF$_3$H, all of the low-lying predicted compounds are computed to be wide gap insulators. Our calculations suggest that other analogous phases with $n>2$ could potentially be metastable at these conditions. For example, exploratory calculations on the LiF$_4$H$_3$ stoichiometry predict a structure that contains bent and asymmetric bifluoride anions, as well as infinite HF chains with nearly equal bond lengths, whose enthalpy lies 21.7~meV/atom above the convex hull.  This studies provides the basis for future work exploring the finite temperature stability of Li-F-H phases under high pressure, with the inclusion of anharmonic effects, which are known to be important for light element systems at these conditions.

\section{Acknowledgement}
E.Z. and T.B. acknowledge the NSF (DMR-1827815), and R.H. acknowledges the NSF (DMR-1933622) for financial support. We thank the U.S. Department of Energy, National Nuclear Security Administration, through the Chicago-DOE Alliance Center under Cooperative Agreement DE-NA0003975.
Calculations were performed at the Center for Computational Research at SUNY Buffalo. \cite{ccr} 
We are grateful to Sebastien Hamel for useful discussions. A.S. notes LLNL-JRNL-817757.

\section{Data Availability}
The data of this study are available from the corresponding author upon request.

%
\end{document}